# Effect of annealing induced polymer substrate attachment on swelling dynamics of ultrathin polymer films

**Mojammel H. Mondal and M. Mukherjee**

Surface Physics Division, Saha Institute of Nuclear Physics, 1/AF, Bidhannagar, Kolkata-64, India

ABSTRACT: The effect of annealing on the dynamical behavior of swelling for ultrathin polyacrylamide films deposited on silicon substrates have been studied using X-ray reflectivity technique. The spin coated polyacrylamide films of similar thicknesses were annealed at various temperatures below and above the glass transition temperature of the polymer. The electron density of the films was found to increase systematically on annealing. The swelling dynamics of the annealed films were found to have systematic dependence on the temperature of annealing. The interaction between the substrate and the polymer molecules was found to play important role in the swelling dynamics of the annealed films unlike our earlier observation with as coated films. The chain segments attached directly to the substrate were believed to have restricted freedom of movements compared to the ones that are at a distance from the substrate and relatively free. Accordingly, the dynamical behavior of swelling was modeled in terms of the combination of a free and a restricted component and was found to be in excellent agreement with the data. The diffusion coefficients corresponding to the restricted polymer segments were an order of magnitude smaller than those of the free segments and the fraction of the same was found to increase with annealing at higher temperatures. The overall reduction of swellability of the films was explained in terms of the increase of density of the films and the segmental attachment to the substrate on annealing.

**Introduction**

In recent years, nanometer scale polymer thin films have drawn tremendous interest due to their technological importance particularly in areas like microelectronics, coatings, biomaterials, and membranes. Thin polymer films are interesting because they often exhibit properties that are different from the corresponding bulk polymers due to entropic effects and energetic interactions arising at the interfaces. The understanding of the mobility of polymer chains near surfaces and interfaces for ultra-thin polymer films in presence of solvent are of technological importance in many areas like emulsion, coating, and adhesion.[1] Knowledge of the bulk polymer properties is not sufficient in this case because the equilibrium structural and dynamical behaviors of the polymer chains close to the substrate or at some interface are quite different due to entropic effects and energetic interactions arising at the interfaces. These interfacial effects lead to changes in chain conformation and mobility that affect the properties of the entire polymer film. The interesting phenomena observed for thin confined polymer films have been documented in various studies, as for example, the thickness and film-substrate interaction dependence of glass transition temperature,[2-6] dewetting kinetics,[7-9] rheological properties[10] or chain mobility close to the substrate.[11,12] The phenomenon of solvent absorption into the pores of a polymeric structure have been exploited by several authors to study different aspects of polymeric material such as diffusion of solvents into the pores,[13-16] pore size distribution,[17] viscoelastic properties[18] etc. Elaborate theoretical[19-25] and experimental[26-28] studies are also available for the dynamics of polymer chains in a polymer melt or through some random medium equivalent to a cross-linked polymer gel network.

Most often, the polymers chosen for the experimental studies were insoluble in water. Systematic observations with ultrathin polymer films of water-soluble polymers were not available in the literature until recently [29-33] though they have broad range of applications such as biomaterials, biosensors, and preservation of foods.[34-38] In this article, we discuss the swelling dynamics of ultrathin films of polyacrylamide (PAM), a water-soluble linear chain homopolymer, as a result of annealing at various temperatures. Four spin coated polymer films of similar thicknesses were annealed at different temperatures. The films were swelled at saturated vapor pressure condition at room temperature while their thickness was monitored with time using X-ray reflectivity technique. The dynamics of swelling was modified as a result of annealing of the films. The dynamical behavior of swelling for the films was found to have systematic dependence on the temperature of annealing. It was found that the polymer-substrate attachment plays very important role in the swelling dynamics of the annealed films.

**Experimental Details**

**Sample Preparation**. High molecular weight ($5 \times 10^6$) polyacrylamide (Supplied by Polysciences Inc., USA) was taken for thin film preparation. Films were prepared on silicon (100) substrate by spin coating method. We have used 4 mg/ml aqueous solution of the polymer for this purpose. During the spinning, clean and warm (60°C) air was flown gently over the sol using a homemade arrangement to facilitate faster evaporation of water. Before coating, silicon wafers were cleaned by RCA cleaning method, where the wafers were boiled at 100°C for about 15 minutes in a solution of H₂O, NH₄OH and H₂O₂ (volume ratio, 2:1:1). The wafers were then thoroughly rinsed with Millipore water. Apart from cleaning, this treatment enhances the hydrophilicity of the silicon surface by introducing –OH dangling bonds on them which help better attachment of the water soluble polymers with the substrate. To study the effect of annealing at various temperatures on the swelling dynamics of the films it was necessary to have films with similar thicknesses, therefore the films were prepared with identical spinning speed from polyacrylamide solution. All the films were prepared at a spinning speed of 800 rpm.

**X-ray Reflectivity.** X-ray reflectivity is one of the best nondestructive methods to measure the thickness and electron density along the depth of polymeric films; here we have used this technique to study the swelling dynamics of the annealed ultrathin polyacrylamide films. X-ray reflectivity data were collected in our laboratory setup with CuKα radiation obtained from copper sealed tube anode (Bruker AXS, D8 Discover) followed by a Göbel mirror. Specular scans with identical incident and scattered angles for X-ray were taken as a function of momentum transfer vector $q_z$ normal to the surface ($q_z = (4\pi/\lambda) \sin\theta$, with $\theta$ equal to the incident and exit angle of the X-ray and $\lambda = 1.54$Å, the wavelength of the radiation).

To release the strain developed into the films during spin coating, all the samples were allowed to swell in saturated vapor environment for two hours. Initial thicknesses of the films were measured by X-ray reflectivity technique while the samples were kept in a cell under rotary vacuum (~5x10⁻² Torr) as described earlier.[29] The films were then annealed below and above the glass transition temperature $T_g$ (165°C) of the polymer at 125°C, 150°C, 175°C and 200°C for 3 hours in vacuum. The films were designated as A, B, C and D respectively. X-ray reflectivity data for all the films were collected at room temperature in vacuum using the cell. To study the swelling of the films a small container of water was inserted into the cell so that the films could swell in the saturated water vapor environment. Swelling causes increase in film thickness and the same was monitored in situ as a function of swelling time using X-ray reflectivity methods. The reflectivity data for the swelling of the films were taken as a function of time. The $q_z$ range for the data collection during swelling was carefully optimized to accommodate sufficient number of thickness oscillations. Reasonably good statistics were obtained in 10-15 min, during which data were collected for each thickness.

To obtain information about the thickness and electron density of the films, the reflectivity data were analyzed using Parrat formalism [39] modified to include interfacial roughness.[40] For the analysis of the X-ray reflectivity data, the input electron density profiles were divided into several boxes of thickness equal or more than the depth resolution ($2\pi/q_{max} \sim 10$Å, where $q_{max}$ is the maximum value of wave vector transfer normal to the surface ($q_z$) for a given reflectivity data) and the interfacial roughness were kept within 2-8 Å. During the analysis, the roughness of the polymer surface, the electron density, the thickness of the films and the roughness of the substrate were used as fitting parameters.

**Results and Discussion**

In table I we have tabulated various parameters for the films. It can be observed from the table that the initial thickness of the films were very similar as they were prepared at same spinning speed. Reduction of thickness for all the films was observed on annealing at various temperatures. Normalised shrinkages in thickness for the films defined as $(R_1-R_2)/R_2$, where $R_1$ and $R_2$ are the film thickness before and after annealing respectively, were found to be within 2.8 to 10.4% and found to have no systematic dependence on the annealing temperatures. The average electron density of the films appears to increase systematically with the increase in the annealing temperature as shown in the last column of the table. This indicates that there was increase of compactness of the films through reduction of average free volume with increase of annealing temperature.

In figure 1 we have plotted the fitted (solid line) specular reflectivity data (symbols) collected at room temperature in vacuum for all the films. In the inset of figure 1 we have shown the corresponding electron density profiles obtained from the fitting of the reflectivity data. The observed electron density profiles show similar nature for all the annealed films except for the one that was annealed at 175°C (film C), which shows lower electron density near polymer substrate interface. An additional large wavelength oscillation in the reflectivity data for the film C unlike other three films also justifies the dip in the electron density profile near the substrate. The lower electron density near the substrate indicates that fewer polymer segments were attached to the substrate in this film which suggests that the attachment of this particular film to the substrate was less compared to the other three films.

To study the swelling of the annealed films, all the films were exposed to saturated vapor pressure of water and the thicknesses of the films were monitored using X-ray reflectivity with time. The thicknesses of the films were found to increase until they reached a saturated value after about 6 hours of swelling. As the thickness of the polymer films in the present experiment are less than the radius of gyration $R_g$ (~100 nm) of the polymer, it may be assumed that the films are constructed from side by side placement of a single layer of segregated polymer coils on the substrate. [43] It can be considered that in the presence of solvent each individual macromolecule swells independently. The diffusion of the polymer chains during swelling or thermal motion of the thin polymer films coated on substrates is observed to occur only along the direction perpendicular to the substrate due to the physical restriction in the other two in-plane dimensions. [29, 41, 42] As discussed earlier, [29] for the analysis of the swelling dynamics of thin polymer films, the end-to-end distance $R(t)$ of a single free chain at time $t$ in one dimension is described with the initial condition $R(t=0) = R_0$ as,

$$R(t) = e^{-(2D/N)t}[R_0^3 + \tfrac{vN^3}{2}(e^{(6D/N)t} - 1)]^{1/3} \qquad (1)$$

where $D$ is the diffusion coefficient of the polymer chains, $N$ is the degree of polymerization and $v$ is the excluded-volume parameter which is a positive quantity for swelling and determines the saturated thickness of the films.

In figure 2 we have plotted the observed thicknesses of the films as a function of swelling time along with the fitted curves obtained using equation 1. It can be observed from the figure that the free chain model described by equation 1 does not agree well with the experimental data except for the film that was annealed at 175°C (film C). The disagreement to the free chain model (equation 1) in figure 2 for the three films indicates that the swelling behavior of the annealed films was likely to be different from that of the unannealed polymer films observed earlier. [29]

The difference in swelling behavior between film C and others may be explained when one takes a closer look to the electron density profiles of the films given in the inset of figure 1. The electron density profile for film C has shown lower density close to the substrate unlike other three films. The agreement of the model (equation 1) with the swelling data for this film along with the observation of lower density close to substrate indicate that the model which describes the swelling of free polymer chains was valid for this particular film, where the attachment of the polymer to the substrate was low. It may be assumed that due to thermal treatment of the films there was enhancement in the number of polymer segments that are attached to the substrate. Therefore in presence of solvent molecules during swelling, these segments would have restricted freedom of movements as compared to the segments that are away from the substrate and relatively free. According to this model, polymer films may be divided into two virtual layers; a first layer consisting of few monomer segments neighboring each point of attachment on the substrate with restricted freedom of segmental motion and a second layer consisting of the rest of the segments between successive points of attachments with higher freedom of segmental movements. Considering this view we model the system as a sum of two components and assume that the dynamics of the two components to be independent. According to this formalism we have modified equation 1 for the description of the dynamics of a "free" and a "restricted" fraction of chains in the films where the end-to-end distance $R(t)$ of the polymer chain is given by,

$$R(t) = R_f(t) + R_r(t) \qquad (2)$$

where,

$$R_f(t) = e^{-(2D_f/N_f)t}[R_f^{\,3} + \tfrac{v_f N_f^{\,3}}{2}(e^{(6D_f/N_f)t} - 1)]^{1/3}$$

and,

$$R_r(t) = e^{-(2D_r/N_r)t}[R_r^{\,3} + \tfrac{v_r N_r^{\,3}}{2}(e^{(6D_r/N_r)t} - 1)]^{1/3}$$

Here $R_f(t)$ and $R_r(t)$ represent the thickness of free and restricted polymer chains respectively. $D_f$ and $D_r$ are the respective diffusion coefficients for the free and the restricted parts. $N$ is the degree of polymerization. The excluded volume parameters for the two components $v_f$ and $v_r$ represent the repulsive interaction responsible for the swelling. $N_f$ and $N_r$ denote the number of segments in the free and the restricted components respectively. According to our model polymer molecules attached to the substrate behave like a chain with alternate occurrence of free and restricted components as shown by the model in figure 3. Although in principle it is possible to determine these numbers from the fitting of our data, the introduction of additional parameters are likely to increase the uncertainty of our analysis. In order to have better confidence in our fitting we have replaced both the parameters by N/2 in equation 2 as a first approximation. $R_f$ and $R_r$ are the initial thicknesses for the free and the attached portions of the film, respectively. It may be noted that these are not true "thicknesses" rather they represent effective lengths of the restricted and free chains in a coil.

In figure 4 we have plotted the observed thicknesses of the films as a function of swelling time along with the fitted curves obtained using equation 2. The data for film C is not shown here as it can be fitted well with equation 1. It may be noted that the excluded-volume parameters and the diffusion coefficients of the free and the restricted parts of the films were kept as floating parameters for the fitting. The parameters $R_f$ and $R_r$ were also allowed to vary within the condition that the total initial thickness $R_0 = R_f + R_r$ remains fixed. Excellent agreement of equation 2 with all the experimental data indicate that the swelling of the films can be described in terms of two independent dynamics corresponding to the two parts of the films as described above. The attachment of the polymer film on the substrate depends on several parameters such as roughness and chemical nature of the substrate. It is not clear at this stage as to why the attachment of film C that was annealed at 175°C was not good although all the films were prepared and annealed under similar conditions.

However, to facilitate better attachment, film C was annealed again at 200°C for one hour. The electron density profile of the dry film shows a shallow dip near the substrate indicating partial improvement in the film substrate attachment after the second annealing. Swelling study of the film was performed again after the annealing. The swelling dynamics of the film was found to be unaltered (data fits well with equation 1) compared to the behavior of the film annealed at 175°C. A third attempt of annealing (200°C for 1 hr) was tried with film C. It was interesting to note that after the third annealing the dip in the electron density profile near the substrate was completely disappeared indicating strong polymer substrate attachment. Study of swelling dynamics was again performed using X-ray reflectivity after this third stage of annealing. The swelling behavior of film C at this stage was found to be similar to the other three films (A, B and D) and the data was fitted with equation 2 as fitting with equation 1 was no longer found to be adequate. In figure 5 X-ray reflectivity data for dry film C after different stages of annealing have been shown. The large wavelength oscillation corresponding to the dip in the electron density profile near the substrate is found to diminish with annealing. The inset of the figure shows corresponding electron density profiles of the film. In figure 6 the thickness change as a function of swelling time for film C have been given for different stages of annealing along with the best fit lines. It can be observed from the figure that the equation 1 gives excellent agreement with the data for the film annealed at 175°C (curve C) and after a second annealing at 200°C (curve C1). The figure also shows the data of the film after a third annealing at 200°C (curve C2). The best fits corresponding to both equations 1 and 2 are shown for curve C2. It can be clearly observed that equation 1 no longer fits well with the data whereas equation 2 was in excellent agreement. From the observation of figure 5 and 6 it was further justified that the swelling dynamics of the annealed polymer films that are attached to the substrate should be represented in terms of swelling of two independent components.

In figure 7 we have shown the two sets of parameters obtained from fitting of equation 2 (figure 4 and 6) as a function of annealing temperature. All the parameters in figure 7 shows two sets of clearly distinct values corresponding to the two independent motions during swelling of the annealed films. Figure 7a shows that the two diffusion coefficients namely slow and fast differ by about one order of magnitude and do not change much with annealing temperature. This indicates that the movements of the polymer segments that are attached to the substrate are significantly slowed down by the attractive interaction of the substrate. Figure 7b shows the two sets of normalized excluded volume parameters (defined as $v/R_0^3$) that determine the saturated thickness of the films. The values are similar at lower annealing temperature. With increase in the annealing temperature the parameter corresponding to the faster diffusing component was found to decrease whereas the other component remains nearly constant. From equation 1 (or 2) it can be observed that $R(t) = (vN^3/2)^{1/3}$ as $t$ tends to infinity. Therefore the saturation thickness of a film is dependent on the product of $v$ and $N^3$. According to our model the number of attachments of a chain increases with annealing. Therefore the average length (or number of segments $N$) between two successive attachments decreases for free segments. The decrease of $v_f$ (calculated for constant $N$) in this case may be attributed to the decrease of $N$ for annealing of the films at higher temperature. On the other hand although the number of attachment increases the length of each restricted segment remains unaltered as they are determined by a fixed distance from the point of attachment upto which the effect of the interaction persists. The corresponding parameter $v_r$ remains nearly constant in this case.

In figure 7c we have plotted the thickness fraction or the effective length of the two components as a function of annealing temperature. The fraction of thickness corresponding to restricted segments was lower at lower annealing temperature and was found to increase with the increase of annealing temperature. This trend is reciprocated by the thickness fraction of the free component. For film A the restricted and free components of thicknesses were about 30% and 70% and at highest temperature of annealing (film D) the numbers were 70% and 30% respectively. The results indicate that with annealing at higher temperature the thickness fraction of the attached part of a polymer chain increase progressively. As the number of attachments of the chains increase, the total length of the restricted segments increases, although the length of each segment remains unaltered and as a result the same for free segments decrease with annealing at higher temperature. This is reflected in the behavior of $R_f$ and $R_r$. In figure 7a and 7b the parameters (using equation 1) corresponding to film C and C1 are shown as asterisks. In figure 7c the parameters corresponding to film C and C1 are not shown as these values are trivially 1. The values corresponding to film C2 are shown as triangles in figure 7. It may be noted that the physical picture depicted by equation 1 and the free component of equation 2 are very similar. It can be seen from figure 7 that the diffusion coefficients obtained for the free components are in good agreement with those obtained from equation 1. Similar matching is obtained for the excluded volume parameters also. As the values of these parameters strongly dependent on the value of $N$ the above agreement shows that the choice of $N_f$ and $N_r$ as *N/2* as a first approximation is not incorrect.

The total change of thickness of the films due to swelling may be measures as swellibility defined as *($R_{max}$-$R_0$)/$R_0$*, where $R_0$ and $R_{max}$ are the initial and the saturated thickness of the films. In figure 8 we have plotted the swellabiliy of all the annealed films as a function of annealing temperature. It was observed from figure 8 that the swellability of the films decreases with annealing temperature. Moreover, the swellability of film C remains unchanged after a second annealing of the film at 200°C for one hour. It may be noted that after this second annealing the attachment of the film to the substrate does not improve very much as can be observed by the dip in the electron density profile in figure 5. After the third annealing at 200°C for one hour, the swellability of the film again decrease when there was better polymer substrate attachment (absence of dip near substrate in electron density profile, figure 5). The results of figure 8 indicate that there are two effects of annealing that are responsible for the reduction of swellability observed here, namely the increase of density of the films and the enhancement of the number of segments attached to the substrate with increase in annealing temperature.

For completeness of the study we have attempted a different model to describe the swelling of the films in terms of an additional component of the free energy corresponding to the polymer-substrate interaction given by $F_s \cong -k_B T N_s \frac{a^2}{R^2}$, which essentially takes into account of interaction between two surfaces (polymer and substrate) at a separation $R$.[43] When this term is added to the Flory free energy, $F(R) = k_B T \left( \frac{R^2}{N} + v \frac{N^2}{R^d} \right)$, the differential equation (equation 1 of reference 29) cannot be solved analytically for $R(t)$. In another approach the film was virtually divided into two layers with a top free layer and a polymer surface consisting of $N_s$ segments on the substrate whose contribution to the swelling due to the surface free energy $F_s$ could be added with the swelling of the top free layer. In this approach the additional component of swelling corresponding to $F_s$ can be described as,

$$R_s(t) \cong (R_s^4 - 8 D_s N_s a^2 t)^{1/4} \qquad (3)$$

Where $D_s$ is the diffusion coefficient of the polymer chains, $R_s$ is the initial thickness of the surface layer and $a$ is the effective monomer length.

It may be noted that the combination of equation 1 and 3 does not represent our swelling data properly. This indicates that the polymer-substrate interaction can not be estimated correctly in terms of the interaction between two surfaces.

**Conclusions**

We have studied the effect of annealing on swelling dynamics of ultrathin polyacrylamide films in the presence of saturated water vapor at room temperature using X-ray reflectivity technique. The films were prepared by spin-coating method on silicon substrate. The films of similar thicknesses were annealed at different temperatures. The swelling dynamics of the annealed films was modified as a result of annealing and was found to have systematic dependence on the temperature of annealing. The attractive polymer substrate interaction was found to play important role in the swelling of the annealed films. Here a fraction of the polymer segments believed to be directly attached to the substrate was found to have restricted freedom of movements due to the attractive polymer substrate interaction. The rest of the polymer segments were assumed to be relatively free. Following the arguments, the change of thickness of the films as a function of time was described as a combination of swelling of a restricted and a free component of the polymer segments. The fraction of the thickness corresponding to the restricted segments was increased with annealing temperature as a result of increased polymer-substrate attachment on annealing. The diffusion coefficients of the restricted polymer segments were about one order of magnitude smaller than those of the free segments. The overall swellability of the films was found to decreases with annealing temperature. The reduction of swellability was attributed to the increase of density of the films and the enhancement of the number of segments attached to the substrate. A separate model to describe the swelling of the annealed films in terms of an additional component of the free energy corresponding to polymer substrate interaction along with the entropic and the excluded volume contributions was found to be inappropriate to describe the present data.

**Acknowledgements**

Authors thankfully acknowledge Prof. S. Hazra for extending the X-ray reflectivity facility for the study.

**References and Notes**

(1) Sanchez, I. C. *Physics of Polymer Surfaces and Interfaces*; Butterworth-Heinemann: Boston, MA, 1992.

(2) Orts, W. J.; van Zanten, J. H.; Wu, W. L.; Satija, S. K. *Phys. Rev. Lett.* **1993**, *71*, 867.

(3) Forrest, J. A.; Dalnoki-Veress, K.; Stevens, J. R.; Dutcher, J. R. *Phys. Rev. Lett.* **1996**, *77*, 2002.

(4) Keddie, J. L.; Jones, R. A. L.; Cory, R. A. *Faraday Discuss. Chem. Soc.* **1994**, *98*, 219.

(5) Wallace, W. E.; van Zanten, J. H.; Wu, W. L. *Phys. Rev. E* **1996**, *52*, R3329.

(6) Aubouy, M. *Phys. Rev. E* **1997**, *56*, 3370.

(7) Zhao, W.; Rafailovich, M. H.; Sokolov, J.; Fetters, L. J.; Plano, R.; Sanyal, M. K.; Sinha, S. K. *Phys. Rev. Lett.* **1993**, *70*, 1453.

(8) Reiter, G. *Macromolecules* **1994**, *27*, 3046.

(9) Kerle, T.; Yerushalmi-Rozen, R.; Klein, J. *Macromolecules* **1998**, *31*, 422.

(10) Hu. H. W.; Granick, S. *Science* **1992**, *258*, 1339.

(11) Lin, E. K.; Wu, W. L.; Satija, S. K. *Macromolecules* **1997**, *30*, 7224.

(12) Lin, E. K.; Kolb, R.; Satija, S. K.; Wu, W. L. *Macromolecules* **1999**, *32*, 3753.

(13) Muller-Buschbaum, P.; Bauer, E.; Maurer, E.; Cubitt, R. *Physica B* **2006**, *385-386*, 703.

(14) Ghiscllini, M.; Quinzi, M.; Giacinti Baschctti, M.; Doghicri, F.; Costa, G.; Sarti, G. C. *Desalination* **2002**, *149*, 441.

(15) Doghieri, F.; Sarti, G. C. *J. Polym. Sci., Part B: Polym. Phys.* **1997**, *35*, 2245.

(16) van Hooy-Corstjeens, C. S. J.; Magusin, P. C. M. M.; Rastogi, S.; Lemstra, P. J. *macromolecules* **2002**, *35*, 6630.

(17) Lee, H. J.; Soles, C. L.; Liu, D. W.; Bauer, B. J.; Wu, W. L. *J. Polym. Sci., Part B: Polym. Phys.* **2002**, *40*, 2170.

(18) Domack, A.; Johannsmann, D. *J. Appl. Phys.* **1996**, *80*, 2599.

(19) Muthukumar, M.; Baumgartner, A. *Macromolecules* **1989**, *22*, 1937.

(20) Muthukumar, M.; Baumgartner, A. *Macromolecules* **1989**, *22*, 1941.

(21) Machta, J. *Phys. Rev. A* **1989**, *40*, 1720.

(22) Ebert, U.; Baumgartner, A.; Schafer, L. *Phys. Rev. E* **1996**, *53*, 950.

(23) Shaffer, J. S. *J. Chem. Phys.* **1994**, *101*, 4205.

(24) Kremer, K.; Grest, G. S. *J. Chem. Phys.* **1990**, *92*, 5057.

(25) Slater, G. W.; Wu, S. Y. *Phys. Rev. Lett.* **1995**, *75*, 164.

(26) Bucknall, D. G.; Butler, S. A.; Higgins, J. S. *Macromolecules* **1999**, *32*, 5453.

(27) Perez-Salas, U.; Briber, R. M.; Hamilton, W. A.; Rafailovich, M. H.; Sokolov, J.; Nasser, L. *Macromolecules* **2002**, *35*, 6638.

(28) Paul, W.; Smith, G. D.; Yoon, D. Y.; Farago, B.; Rathgeber, S.; Zirkel, A.; Willner, L.; Richter, D. *Phys. Rev. Lett.* **1998**, *80*, 2346.

(29) Singh, Amarjeet.; Mukherjee, M. *Macromolecules* **2003**, *36*, 8728.

(30) Singh, Amarjeet.; Mukherjee, M. *Macromolecules* **2005**, *38*, 8795.

(31) Singh, Amarjeet.; Mukherjee, M. *Phys. ReV. E* **2004**, *70*, 051608.

(32) Mukherjee, M.; Singh, Amarjeet.; Daillant,J.; Alain, M. and Cousin, F. *Macromolecules* **2007**, *40,* 1073-1080

(33) Mukherjee, M.; Singh, A. *phys. Stat. sol. (b)* **2007**, *244*, 928.

(34) Fair, B. D.; Jamieson, A. M. *J. Colloid Interface Sci.* **1980**, *77*, 525.

(35) Chan, B. M. C.; Brash, J. L. *J. Colloid Interface Sci.* **1984**, *82*, 217.

(36) Arnebrant, T.; Nylander, T. *J. Colloid Interface Sci.* **1986**, *111*, 529.

(37) Rames, A.; Williums, D. F. *Biomaterials* **1992**, *13*, 731.

(38) Green, R. J.; Hopkinson, I.; Jones, R. A. L. *Langmuir* **1999**, *15*, 5102.

(39) Parratt, L. G. *Phys. Rev.* **1954**, *95*, 359.

(40) Russel, T. P. *Materials Science Reports*, vol. 5, Elsevier Science Publ., North-Holland (1990)

(41) Mukherjee, M.; Bhattacharya, M.; Sanyal, M. K.; Geue, Th.; Grenzer, J.; Pietsch, U. *Phys. Rev. E* **2002**, *66*, 061801.

(42) DeMagio, G. B.; Frieze, W. E.; Gidley, D. W.; Zhu, M.; Haristov, H. A.; Yee, A. F. *Phys. Rev. Lett.* **1996**, *78*, 1524.

(43) de Gennes, P. G. *Scaling Concepts in Polymer Physics*; Cornell University Press: Ithaca, NY, 1979.

Table I

Various Parameters of the Polyacrylamide Films before and after Annealing.

| annealing temperature (°C) | film thickness at room temperature (Å) $R_1$ | film thickness after annealing (Å) $R_2$ | normalised shrinkage (%) $((R_1-R_2)/R_2)*100$ | average electron density (Å$^{-3}$) |
|---|---|---|---|---|
| 125±1.0 | 618 | 560 | 10.3 | 0.41±0.005 |
| 150±1.0 | 620 | 603 | 2.8 | 0.42±0.005 |
| 175±1.0 | 604 | 580<br><br>573 [a]<br><br>564 [b] | 4.1 | 0.43±0.005<br><br>0.43±0.005 [a]<br><br>0.43±0.005 [b] |
| 200±1.0 | 603 | 546 | 10.4 | 0.44±0.005 |

a - total annealing, 175°C for 3 hr + 200°C for 1 hr.

b - total annealing, 175°C for 3 hr + 200°C for 2 hr.

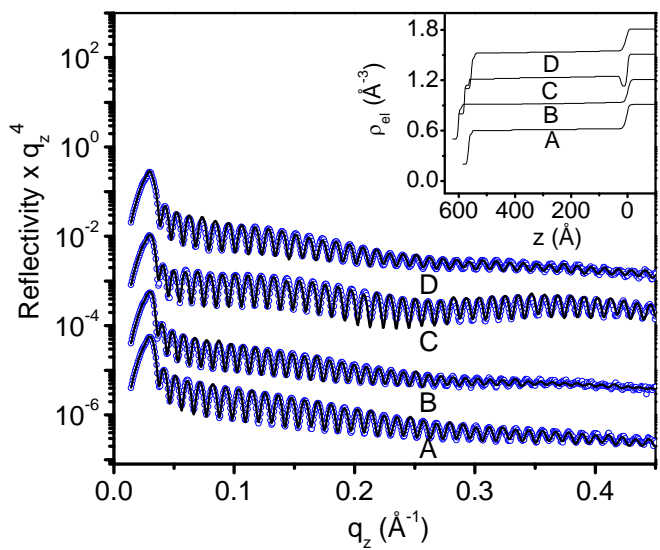

**Figure 1.** X-ray reflectivity data is shown as symbols with fittings as lines for the films A, B, C, D that were annealed at 125°C, 150°C, 175°C, 200°C respectively. The inset shows corresponding electron density profiles. The data for both figures have been shifted suitably for clarity.

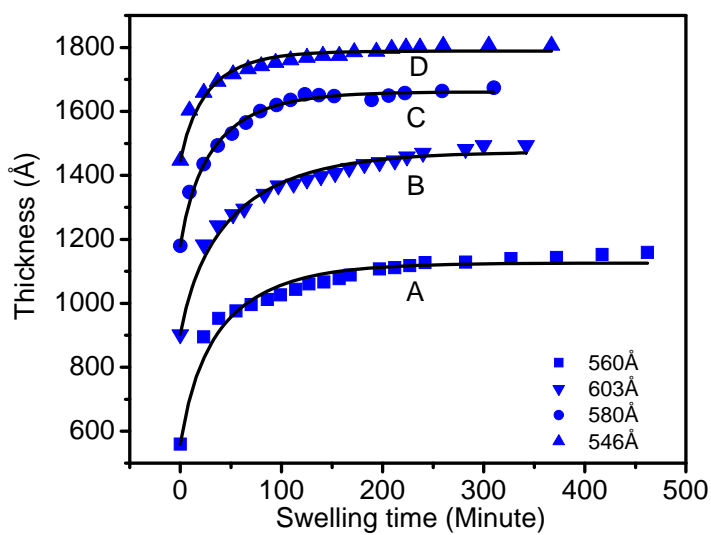

**Figure 2.** Thickness of the annealed films with time as they are exposed to saturated water vapor. A, B, C and D represent the data (symbol) and fit (line) corresponding to the films annealed at 125°C, 150°C, 175°C and 200°C respectively. The lines are obtained by fitting the data with equation 1. Correct fitting is obtained for film C only. Initial thicknesses of the films are shown against the corresponding symbols. The data B, C and D are shifted upward for clarity.

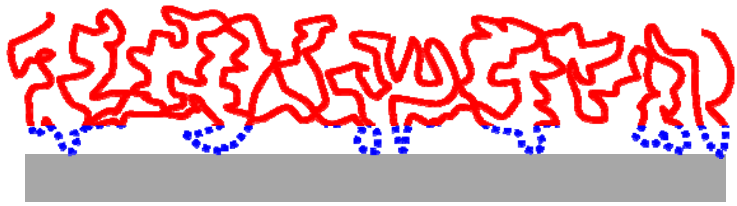

(a)

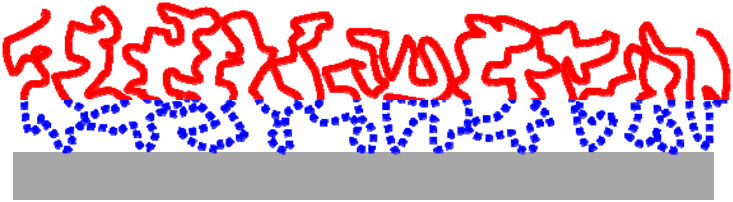

(b)

**Figure 3.** Schematic model of polymer films on substrate for annealing at lower and higher temperature. The solid and the dotted lines represent free and restricted segments respectively. (a) less attachment to the substrate for annealing at lower temperature. (b) more attachment points for annealing at higher temperature.

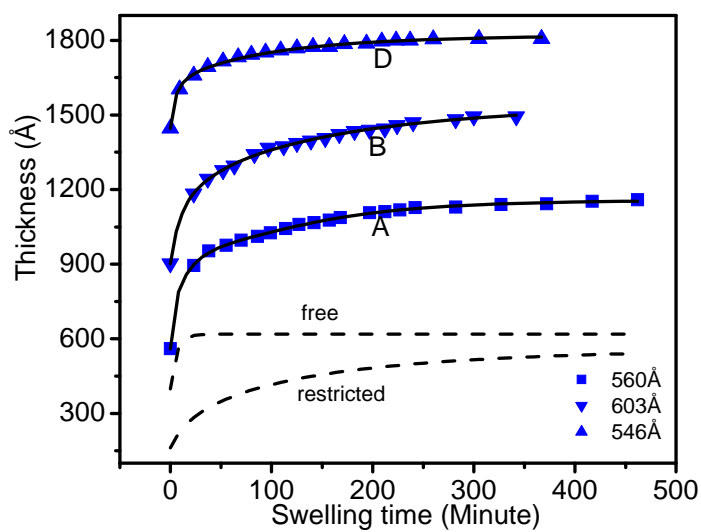

**Figure 4.** Thickness of the annealed films as a function of time as they are exposed to saturated solvent vapor. A, B and D represent the data (symbol) and fit (line) corresponding to the films annealed at 125°C, 150°C and 200°C respectively. The lines are obtained by fitting of the data with equation 2. Initial thicknesses of the films are shown against the corresponding symbols. The films annealed at 150°C and 200°C with thicknesses 603 and 546Å respectively are shifted upward by 300 and 900 Å for clarity. The contribution of the individual components for the fitting of film A is shown by dashed lines.

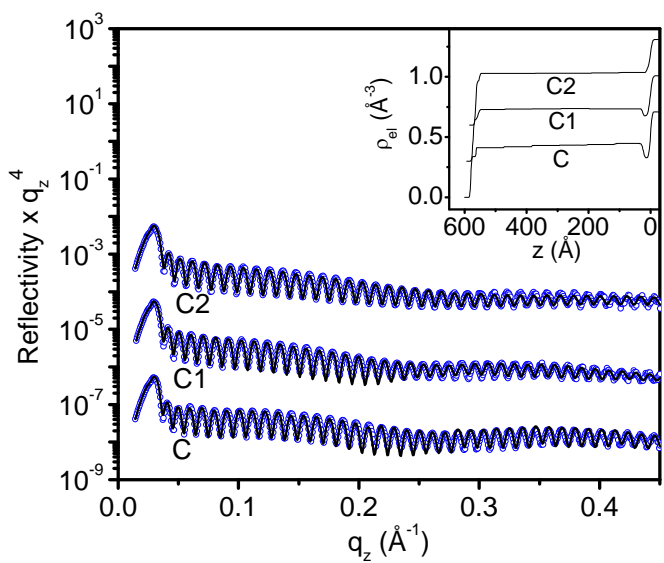

**Figure 5.** X-ray reflectivity data (symbols) along with the fittings (solid line) for polyacrylamide film C that was annealed at 175°C for 3 hours. C1 and C2 represent the reflectivity data (symbol) along with the fittings (solid line) of the same film after annealing at 200°C for 1 and 2 hours respectively. The inset shows the electron density profiles for the film as obtained from the fitting of the reflectivity data. The reflectivity data and the electron density profiles for the film C1 and C2 have been shifted suitably for clarity.

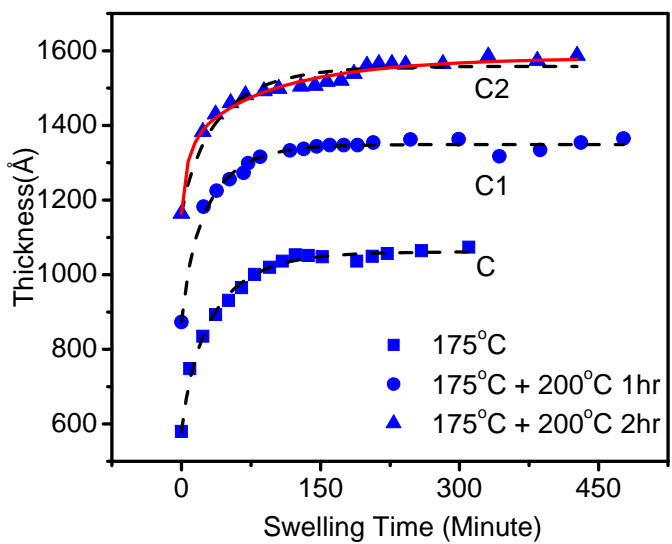

**Figure 6.** Plot of film thickness with swelling time as the film was exposed to saturated solvent vapor. C represents the data (symbol) and fit (line) for the film annealed at 175°C for 3 hours. C1 and C2 represent the data and fit for the same film after annealing at 200°C for 1 and 2 hours respectively. The lines are obtained by fitting the data with equation 1 (dashed) and equation 2 (solid). The data C1 and C2 with initial thicknesses 573Å and 564Å are shifted upward by 300Å and 600Å respectively for clarity.

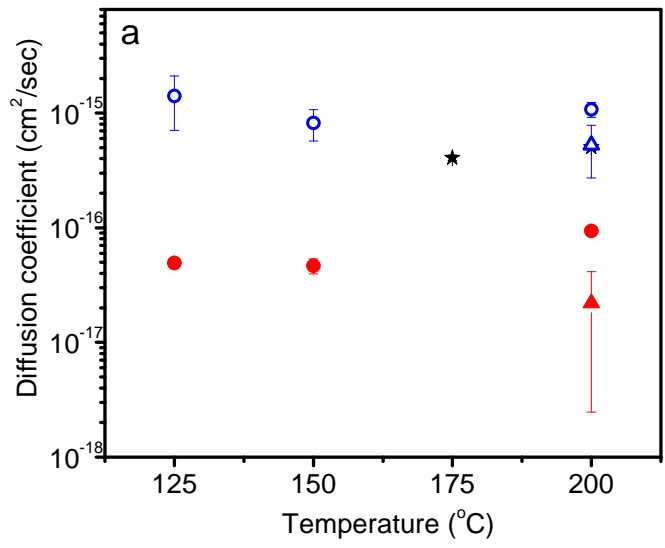

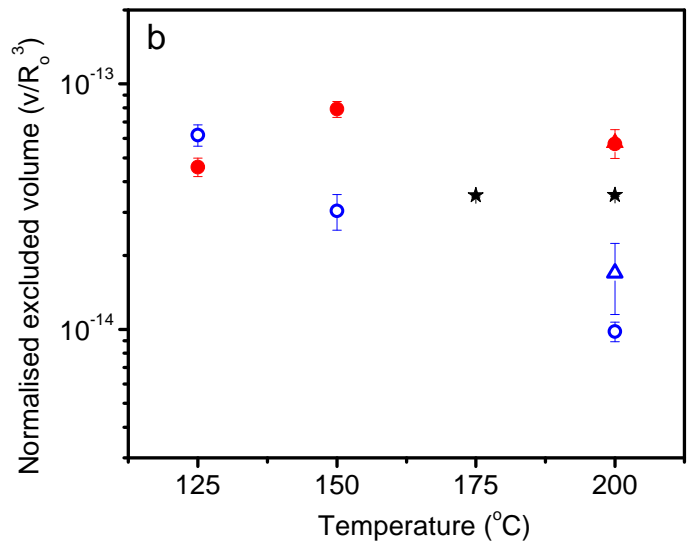

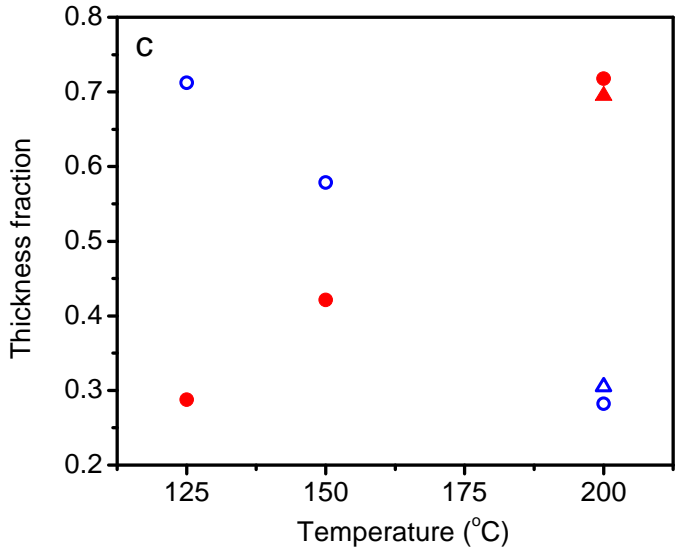

**Figure 7.** Plot of the parameters obtained from fitting of the X-ray reflectivity measurements with swelling time by equation 2 (figure 3 and 5) as a function of temperature.

(a) Diffusion coefficients for the free chain fraction $D_f$ (open symbol) and the restricted chain fraction $D_r$ (solid symbol) of the films as a function of the annealing temperature. The asterisks represent the values for the film C and C1. The solid and open triangles represent the diffusion coefficients for the restricted and free chain fraction of the film that was annealed twice at 200°C (film C2).

(b) The normalized excluded-volume parameters for the free chain fraction $v_f$ (open symbol) and the restricted chain fraction $v_r$ (solid symbol) of the films as a function of annealing temperature. The asterisks represent the values for the film C and C1. The solid and open triangles represent the normalized excluded-volume parameters for the restricted and free chain fraction of the film that was annealed twice at 200°C (film C2).

(c) The normalized thickness for the free chain fraction $R_f$ (open symbol) and the restricted chain fraction $R_r$ (solid symbol) of the films as a function of annealing temperature. The solid and open triangles represent the normalized thicknesses for the restricted and free chain fraction of the film that was annealed twice at 200°C (film C2). The values for film C1 and C2 are not shown in the figure as the values are unity.

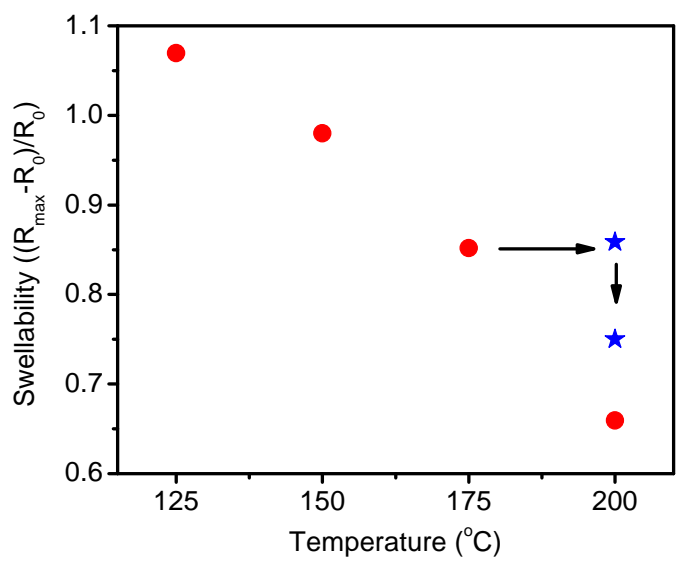

**Figure 8.** Swellability of the annealed films as a function of annealing temperature. Asterisks show the swellability of film C after additional annealing at 200°C. The arrows indicate the direction of change after 1 hr and 2 hrs of annealing at 200°C.

For Table of Contents use only

Title: Effect of annealing induced polymer substrate attachment on swelling dynamics of ultrathin polymer films

Authors: Mojammel H. Mondal and M. Mukherjee

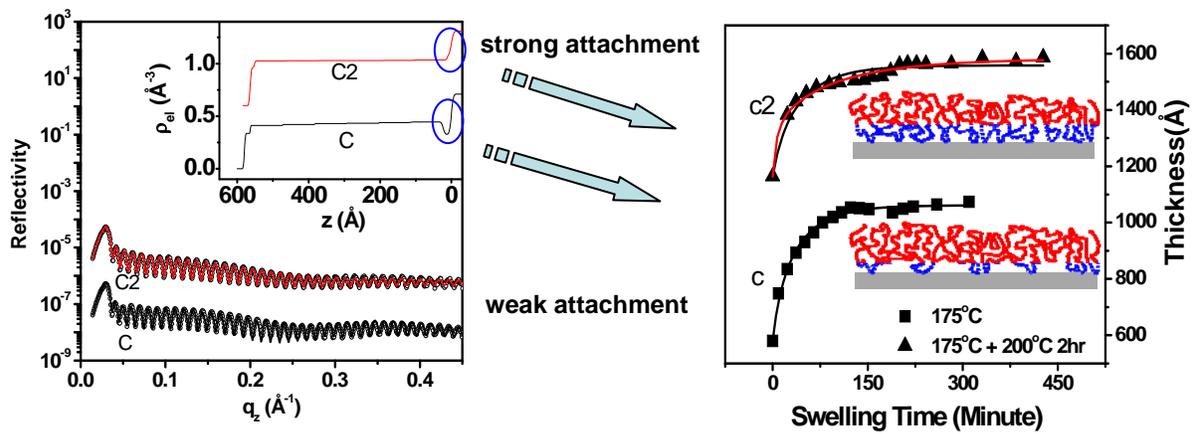